\documentclass[superscriptaddress,aps, prb, twocolumn,showpacs,showkeys,notitlepage,amsmath, amssymb]{revtex4-2}
\usepackage[utf8]{inputenc}
\usepackage[varg]{txfonts}
\usepackage{bm}
\usepackage{soul}
\usepackage{pbox} 
\usepackage[svgnames]{xcolor} 
\usepackage{lipsum}
\usepackage{cancel}
\usepackage{braket}
\usepackage{xspace}
\usepackage{physics}
\usepackage{siunitx}
\usepackage{comment}
\usepackage{upgreek}
\usepackage{colortbl}
\usepackage{graphicx}
\usepackage{rotating}
\usepackage{nicefrac}
\usepackage{multirow}
\usepackage{orcidlink}
\usepackage{longtable}
\usepackage{threeparttable}
\usepackage[normalem]{ulem}
\usepackage[caption=false]{subfig}
\usepackage[referable]{threeparttablex}
\usepackage{booktabs,microtype,afterpage}

\usepackage{hyperref}
\hypersetup{
    colorlinks = true,
    linkcolor = Blue,
    citecolor = Blue,
    urlcolor  = Blue
}
\def \tbsq {Tb-SQ}
\def \tbtrp {Tb-Trp}

\def \muon {$\mu$SR}

\def \abbrref {Ref.}
\def \abbrtab {Tab.}
\def \abbrfig {Fig.}
\def \abbreqn {Eq.}
\def \tbIII {Tb$^{3+}$}
\def \tbIII {Terbium(III)}
\graphicspath{{figures/}} 
\DeclareGraphicsExtensions{.pdf,.PDF,png,.PNG}
\begin{document}

 
\title{\texorpdfstring{\muon}{} evidence of a marked exchange interaction effect on the local spin dynamics of Tb-based molecular nanomagnets}

%
\author{\hspace{1mm}Muhammad Maikudi Isah\orcidlink{0000-0002-6615-2977}}
\affiliation{
Dipartimento di Fisica e Astronomia ``A. Righi'', Universit\'a di Bologna and INFN Sezione di Bologna, I-40127 Bologna, Italy 
}
\author{\hspace{1mm} Lorenzo Sorace\orcidlink{0000-0003-4785-1331}}
\email{lorenzo.sorace@unifi.it}
\affiliation{
Dipartimento di Chimica ``U. Schiff'', Universit\'a di Firenze, Via della Lastruccia 3, and INFN Sezione di Firenze, I-50019 Sesto Fiorentino, Italy}
\author{\hspace{1mm} Alessandro Lascialfari\orcidlink{0000-0002-1687-0436}}
\affiliation{
Dipartimento di Fisica, Universit\'a di Pavia, and INFN Sezione di Pavia, Via Bassi 6, I-27100 Pavia, Italy }
\author{\hspace{1mm} Paolo Arosio\orcidlink{0000-0003-2388-0402}}
\affiliation{
Dipartimento di Fisica ``A. Pontremoli'', Universit\'a degli Studi di Milano, and INFN Sezione di Milano, 20133 Milano, Italy}
\author{\hspace{1mm} Zaher Salman\orcidlink{0000-0002-3431-8135}}
\affiliation{
Laboratory for Muon Spin Spectroscopy, Paul Scherrer Institute, CH-5232 Villigen PSI, Switzerland}
\author{\hspace{1mm} André Moreira Nogueira\orcidlink{0000-0001-9146-3243}}
\affiliation{
Instituto de Química, Universidade Federal do Rio de Janeiro, Rio de Janeiro, RJ, Brazil
}
%
%
\author{\hspace{1mm} Giordano Poneti\orcidlink{0000-0002-1712-4611}}
\affiliation{
Dipartimento di Scienze Ecologiche e Biologiche, Universit\'a della Tuscia, Largo dell'Universit\'a, snc, 01100 Viterbo VT, Italy}
\author{\hspace{1mm} Manuel Mariani}
\email{manuel.mariani@unipv.it}
\affiliation{
Dipartimento di Fisica, Universit\'a di Pavia, Via Bassi 6, I-27100 Pavia, and INFN Sezione di Milano, 20133 Milano, Italy }
\author{\hspace{1mm}Samuele Sanna\orcidlink{0000-0002-4077-5076}}
\affiliation{
Dipartimento di Fisica e Astronomia ``A. Righi'', Universit\'a di Bologna and INFN Sezione di Bologna, I-40127 Bologna, Italy 
}

\date{\today}

%
\begin{abstract}
We report on the spin dynamics of two Terbium-based molecular nanomagnets, namely Tb(DTBSQ)(HBPz$_3$)$_2$ (in short \tbsq{}) and Tb(Trp)(HBPz$_3$)$_2$ (in short \tbtrp{}), investigated by means of longitudinal muon spin relaxation (\muon) measurements as a function of applied field, flanked by AC susceptibility characterization. In the two molecules \tbIII{} magnetic ion has an isostructural coordination sphere, but in the former the \tbIII{}  is coordinated by an organic paramagnetic ligand (3,5 ditertbutylsemiquinonate, SQ), while the latter is coordinated by a diamagnetic one (tropolonate, Trp). Thus \tbsq{} presents an exchange interaction between the \tbIII{} ion and a radical while \tbtrp{} does not. Both the samples exhibit a muon spin-lattice relaxation rate $\lambda_1(T, B_L)$ peak in the temperature range 10-25 K at all applied longitudinal magnetic fields $B_L = 50, 150, 300$ milli-Tesla (mT). In \tbsq{}, $\lambda_1(T, B_L)$ displays a BPP-like behavior led by three different correlations times: the first, dominating for $T\geq 15K$, follows a thermally activated law $\tau_c =\tau_0 \exp(\sigma_A / k_B T)$ with energy barrier $\sigma_A/k_B$, while the second and third ones, dominating respectively for $8<T<15$ K and $T<8$ K, follow a power-law-like behavior $\tau_c = c_0 T^{-\alpha}$ with two different values of $c_0$ and $\alpha$. On the other hand, the temperature and field behavior of $\lambda_1(T, B_L)$ in \tbtrp{} strongly deviates from a BPP law, displaying a strongly anomalous character. Our results indicate that, in the absence of an exchange interaction and maintaining all the other relevant interactions constants, the local spin dynamics of single ion magnets strongly differ from the one observed in the presence of such interaction. The combination of \muon{} and AC susceptibility measurements allows us to disentangle the different Orbach, Raman and direct mechanisms which are the key ingredients that control the spin dynamics in \tbsq{}, and evidence the potentiality of \muon{} in elucidating complex spin dynamics.

\end{abstract}



\maketitle


\section{\label{sec:intro} Introduction}
Molecular nanomagnets (MNMs) are transition metal or lanthanide-based molecules that are made of identical, almost non-interacting clusters, constituted of one or several interacting magnetic ions surrounded by an organic shell \cite{gatteschi2006molecular}. Among them, single-molecule magnets (SMMs) \cite{Christou_Gatteschi_Hendrickson_Sessoli_2000} and single-ion magnets (SIMs) \cite{doi:10.1021/ja029629n} show slow relaxation of the magnetization of pure molecular origin below a given blocking temperature ($T_B$) and are thus promising candidates for high-density information storage \cite{Gao2015, Mannini2010, https://doi.org/10.1002/adma.201506305}, and molecular spintronics \cite{Cornia2017, Rocha2005, Bogani2008} applications. This peculiar behavior stems from the presence of a magnetic anisotropy barrier $\sigma_{\mathrm{A}}$ so that, in view of their exploitation, the possibility to tune $\sigma_{\mathrm{A}}$ via the modification of the ligand shell and/or of the local environment around the magnetic ions is very attractive. From this viewpoint, lanthanide-(Ln)-based SIMs, characterised by large magnetic anisotropy and high magnetic moments, and thus by a high energy barrier $\sigma_{\mathrm{A}}/k_B$, appear very promising, and indeed provided examples of $T_B$ close to liquid N$_2$ temperature \cite{Goodwin2017, https://doi.org/10.1002/anie.201705426, doi:10.1126/science.aav0652, C8SC03907K, Gould2019}. A further advantage of these systems is the possibility to engineer the coupling to a second magnetic center, which provides an additional handle to control the dynamics. It has been shown that a weak exchange interaction strongly influences the secondary magnetization relaxation pathways of SMMs e. g. by minimizing the effect of Quantum Tunneling of Magnetization (QTM) at zero-field, in analogy to exchange bias in permanent magnets that allows to modulate the value of the zero-field magnetic remanence \cite{Habib2011, Zhenhua2024}. However, when measured under an applied field  the observed relaxation  is often faster for these systems due to the state mixing induced by the exchange interaction \cite{doi:10.1021/ja109706y}. The most interesting results have been then obtained when stronger interactions $\textrm{--}$ not at all obvious given the inner nature of the $4f$ orbitals $\textrm{--}$ yield coupled systems behaving as single giant spins with huge anisotropy, resulting in extremely slow magnetization relaxation \cite{doi:10.1126/science.abl5470, doi:10.1021/acs.inorgchem.1c00647, Rinehart2011}. 

The analysis of the magnetization dynamics of these systems has further shown that, beyond the overbarrier process (Orbach) and the underbarrier one (QTM), additional magnetization relaxation pathways play a crucial role \cite{trensal1}.  Of particular relevance are, e.g., direct and Raman processes, each of which shows a specific field and temperature dependence  \cite{C5CS00222B}. It is, thus, of fundamental importance to discriminate the role that each kind of interaction (exchange and spin-orbit coupling, crystal-field effects), spanning several orders of magnitude on the energy scale, plays in driving the relaxation dynamics \cite{GATTESCHI201691}: only a multi-technique approach is able to assure an appropriate investigation on different timescales of suitably designed molecular systems. Despite this, magnetization dynamics in these systems is most often investigated only by using ac susceptometry, which probes a restricted range of frequencies ($0.1~\text{Hz}< \nu <10~\text{kHz}$). On the other hand, local spectroscopic techniques such as muon spin relaxation (\muon{}), which have been proved to be a useful and powerful probe of the spin dynamics in $3d$ polynuclear complexes \cite{Borsa2006}, appear still underused for Ln-based SIMs \cite{PhysRevB.79.220404, C8CC04703K, PhysRevB.97.144414}, despite some of the few reported studies proved interesting and unexpected magnetization dynamics \cite{PhysRevB.100.174416}.

\begin{figure}[]
    \centering
    \includegraphics[width=\columnwidth]{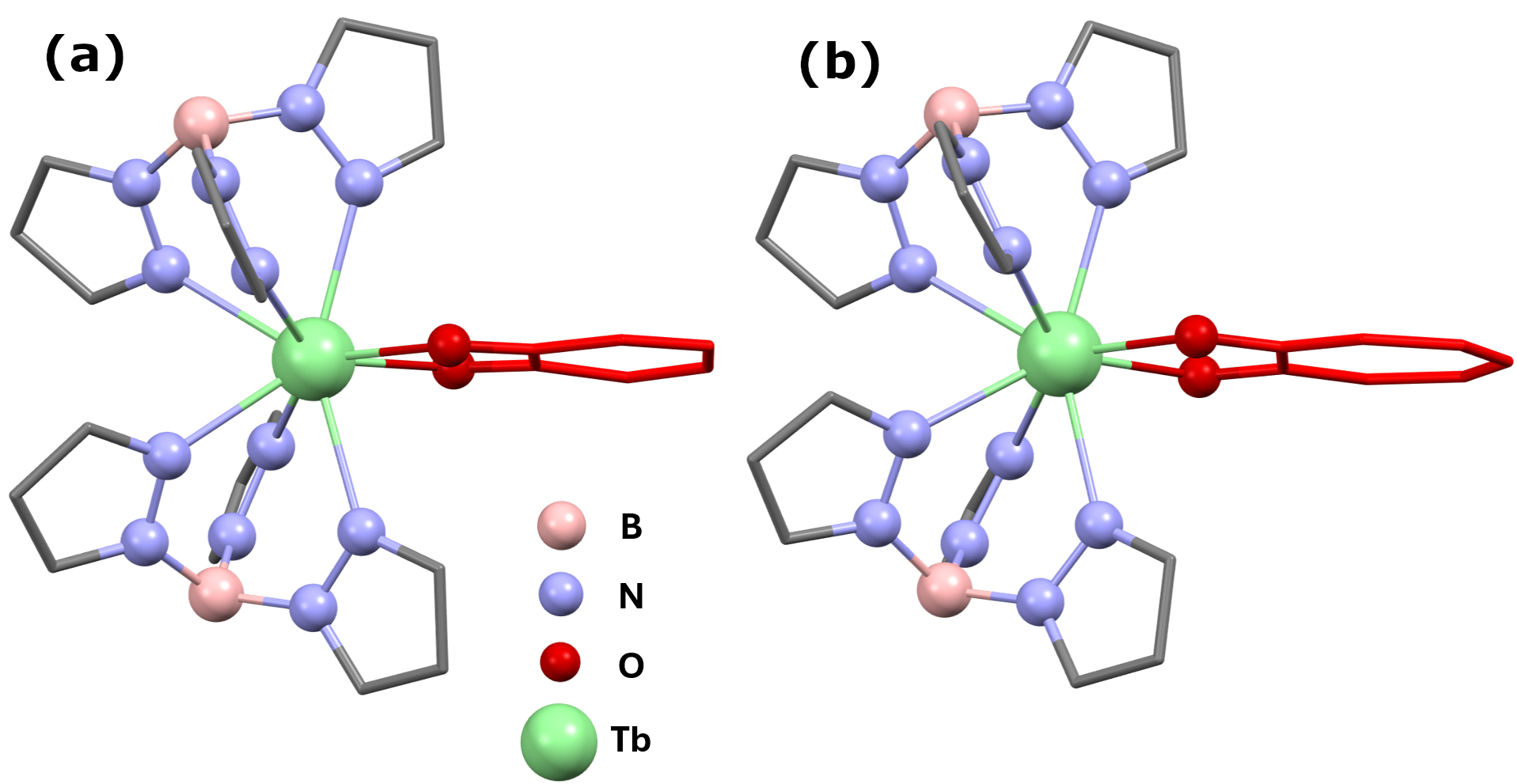}
    \caption{
    {Molecular structures of (a) \tbsq{} and (b) \tbtrp{} evidencing the nearly identical coordination environment of Terbium in the two compounds. Carbon atoms are represented in stick mode for the sake of clarity. For the same reason hydrogen atoms are omitted and the two di-tert-butyl groups in (a) are not shown. SQ is based on a mononegative, six-membered Carbon ring, with radical character, while Trp is based on a mononegative, seven-membered Carbon ring, closed shell. To highlight the close similarity of the two ligands, their carbon skeleton is represented in red.
    }
    }
    \label{fig:TbSQTbTrp_molecular_structure}
\end{figure}

Here, we have investigated spin dynamics of Tb(DTBSQ)(HBPz$_3$)$_2$ (in short \tbsq{}) and Tb(Trp)(HBPz$_3$)$_2$ (in short \tbtrp{}) (DTBSQ=3,5-di-\textit{tert}-butylsemiquinonato, Trp=tropolonate, HBPz$_3$=hydrotrispyrazolyl-borate), by means of longitudinal field muon spin relaxation (LF-\muon{}) aimed at highlighting the difference in the temperature spin dynamics among the two samples, at different applied longitudinal fields. The two complexes are isostructural in their first coordination sphere (see \abbrfig{} \ref{fig:TbSQTbTrp_molecular_structure}) and thus the splitting of the energy level of the lanthanide ion induced  by ligand field is expected to be remarkably similar. However, the semiquinonate (SQ) ligand is paramagnetic and thus gives a Ln-radical exchange interaction, while the tropolonate (Trp) ligand is diamagnetic \cite{B401144A}. Our results show that at intermediate and low temperatures the absence of the radical spin, and consequently of the exchange interaction, modifies heavily the spin dynamics, which strongly departs from the Bloembergen-Purcell-Pound (BPP) law \cite{BLOEMBERGEN1947, PhysRev.73.679}, commonly followed by MNMs. On the other hand, the dynamics observed for \tbsq{} follows a BPP-like law and allows to derive information on the different processes driving the spin dynamics, thus giving hints on the relative importance of exchange and ligand field effects in determining it.



\section{Experimental details}
\label{sec:exp_details}

\subsection{S\MakeLowercase{ample preparation}}
Samples were prepared as microcrystalline powders following the synthetic route reported in \abbrref~\cite{DEI2001135}. Consistency with expected product was checked by Powder X-ray Diffraction, which provided patterns perfectly compatible with the theoretically calculated ones (see \abbrfig{} S1 in the Supplemental Material~\footnote{See Supplemental Material at [URL will be inserted by publisher] for PXRD spectra and additional AC susceptibility data}).

\subsection{AC \MakeLowercase{susceptibility}}
Alternating current (AC) magnetic susceptibility data were collected as a function of field, frequency and temperature by using both a commercial Quantum Design superconducting quantum interferometer device (SQUID) MPMS-XL7 magnetometer, for measurements in the frequency range $0.1~\text{Hz}< \nu <1~\text{kHz}$, and the ACMS inset of the Quantum Design Physical Properties Measurement System (PPMS), for measurements in the frequency range $10~\text{Hz}< \nu <10~\text{kHz}$.

\subsection{M\MakeLowercase{uon spin experiments}}
$\mu$SR experiments were performed on powder samples of both materials at the Swiss Muon Source (S$\mu$S), Paul Scherrer Institut (PSI), Switzerland, using the General-Purpose Surface-Muon (GPS) instrument. In both samples, $\mu$SR spectra were collected in temperature range $1.5< T <200~\text{K}$ at three different applied fields $B_L$ = 50, 150, and 300 mT, parallel to the initial muon spin polarization, namely in the longitudinal field configuration.
Data at constant temperature $T\simeq1.6$ K for \tbsq{} and $T\simeq1.5$ K for \tbtrp{} were also collected as a function of $B_L$ in the range $5< B_L <750~\text{mT}$.

In a $\mu$SR experiment \cite{Schenck1985,  doi:10.1080/001075199181521, AYaouanc2011, BlundellmuSR2020}, 100\%-spin polarized positive muons with the polarization aligned antiparallel to their momentum are implanted into the sample. The muon spin-$\frac{1}{2}$ acts as a local magnetic probe that experiencies a spin precession around an internal magnetic field $B_{\mu}$ at the muon stopping site(s), with a Larmor (angular) frequency $\omega_L^{\mu} = \gamma^{\mu} B^{\mu}$, where $\gamma^{\mu} = 2\pi \times 135.5~\mathrm{MHz/Tesla}$ is the muon gyromagnetic ratio.
The positive muon,  which stops at interstitial sites \cite{PhysRevB.87.115148, Möller_2013, doi:10.1021/jp5125876, PhysRevB.97.174414, PhysRevMaterials.5.044411}, decays with a mean timelife of about 2.2 $\mu$s
emitting a positron ($e^+$) preferentially along the muon spin direction \cite{PhysRev.105.1671}. 
The positrons are detected by using detectors placed around the sample in the forward (F) and backward (B) directions, established with respect to the initial one of the muon spin polarization. The observed quantity is the decay positron asymmetry function, $A(t)$, measured as a function of time, defined as the difference of the number of positrons detected in the $N_F$ and $N_B$ detectors, normalized by their sum $A(t) =  \big[N_F(t)-\alpha N_B(t)\big]/\big[N_F(t)+\alpha N_B(t)\big]$, where $\alpha$ is an experimental calibration constant which differs from unity due to non-uniform detector efficiency. 
The asymmetry is proportional to the average muon spin polarization function $\mathcal{P}(t)$ which contains the information (static or quasi-static and/or dynamic) about the local internal magnetic field and the muon spin-spin and spin-lattice relaxation at the stopping site(s): $A(t) =  A_0\mathcal{P}(t)$, where $A_0$ is the initial asymmetry which is instrument dependent.
All the \muon{} data were analyzed using the PSI fitting program \cite{SUTER201269}, MUSRFIT.


%
%
%
%
%
%
\section{RESULTS AND DISCUSSION}
\label{sec:results_and_discuss}

\subsection{AC susceptibility}
\label{sec:acsusc}
\begin{figure}[ht]
    \includegraphics[width=1.0\linewidth]{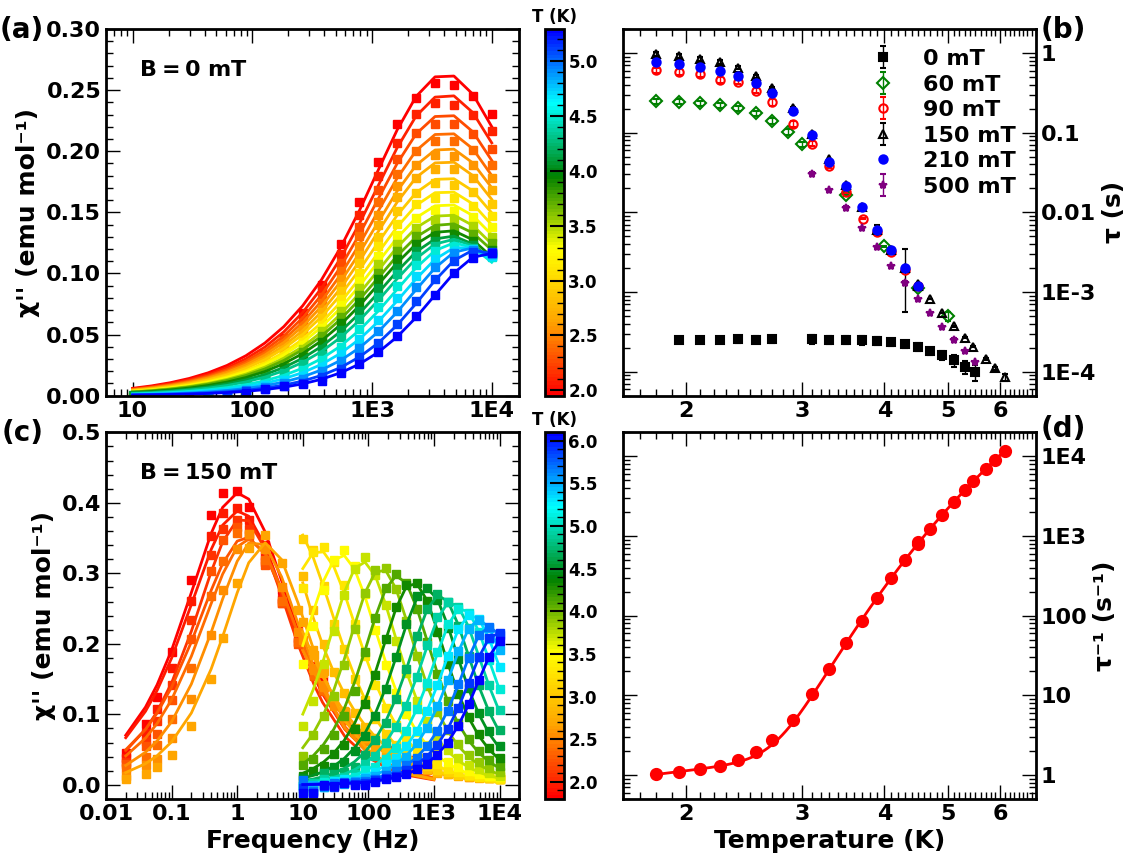}
    \caption{AC susceptibility data for \tbsq{}. (a) Frequency dependence of the imaginary component of the susceptibility measured as a function of temperature in zero field. Continuous lines represent the best fit using a generalized Debye model \cite{gatteschi2006molecular}. (b) Log-log plot of the temperature dependence of the magnetization relaxation time at different fields obtained from fits of AC susceptibility data. (c) Frequency dependence of the imaginary component of the susceptibility measured as a function of temperature at $B_{\mathrm{}}=150~\text{mT}$, providing the slowest relaxation. Continuous lines represent the best fit using a generalized Debye model \cite{gatteschi2006molecular}. (d) Log-log plot of the temperature dependence of the magnetization relaxation rate measured in an external field of $150~\text{mT}$; the continuous line represent the fit obtained with parameters reported in the text. 
    }
    \label{fig:acpanel}
\end{figure}

While direct current (DC) magnetic characterization had already been discussed for both \tbtrp{} and \tbsq{} in \cite{B401144A}, no ac susceptibility data on the two complexes can be found in literature. In zero field \tbtrp{} at low temperature (2 K) does not show any significant slow dynamics in the 10 Hz -  10 KHz range. On applying a DC field $B_{\mathrm{DC}} =150~\text{mT}$, two maxima are clearly observed at the two extremes of the experimental window. On further increasing the field above $B_{\mathrm{DC}} =150~\text{mT}$ the fastest process accelerates and essentially becomes unobservable; on increasing temperature the slow process is strongly suppressed (see \abbrfig{} S2 and S3 in the Supplemental Material). Given the extremely fast dynamics a reliable determination of the characteristics relaxation times as a function of field and temperature is not possible. The dynamics observed on TbSQ is completely different. First and foremost, slow relaxation is clearly observed in zero field (\abbrfig{} \ref{fig:acpanel}), with rates which are not much dependent on the temperature. {The relaxation time in zero field extracted by the fit \cite{fitacprogram} of the $\chi^{\prime}$ and $\chi^{\prime\prime}$ curves by a generalized Debye function \cite{gatteschi2006molecular} is temperature independent up to 4 K, thus clearly indicating a dominant QTM process (\abbrfig{} \ref{fig:acpanel}(b)) in this temperature region. On increasing field, the relaxation becomes slower and reaches a minimum at 150 mT, before accelerating again on further increasing the field (\abbrfig{} \ref{fig:acpanel}(c)). Such a behavior can be safely attributed to the competing effects of the QTM process (which is slowed down by the application of a field \cite{trensal1}) and the direct one (which scales with the field as the fourth power \cite{trensal1}). The temperature dependence of the relaxation rate in an applied field of 150 mT (\abbrfig{} \ref{fig:acpanel}(b)) measured up to 6 K, allows us to determine the contribution of the different thermally activated processes to the relaxation.} A log-log plot clearly evidences the deviation of the thermally activated behaviour from a simple Raman process, which should be linear. Accordingly, the temperature dependence of the relaxation rate could be fitted  (\abbrfig{} \ref{fig:acpanel}(d)) using the following equation:

\begin{eqnarray}
    \label{eqn:Fit_ac}
   \tau^{-1}  &=&  DT + \tau_0^{-1} \exp(-\sigma_{\mathrm{A}}/k_BT)
\end{eqnarray}

\noindent
where the first term represents the direct process and the second one the Orbach process. The best-fit curve was obtained by using the following parameters: $D = 0.525(1) ~\text{s}^{-1}\text{K}^{-1}$, $\tau_0^{-1}  = 2.21(3) \times 10^7 ~\text{s}^{-1}$ and $\sigma_{\mathrm{A}}/k_B = 46.0(1) ~\text{K}$. 
\begin{figure*}[t]
    \centering
    \includegraphics[width = 1.0\linewidth]{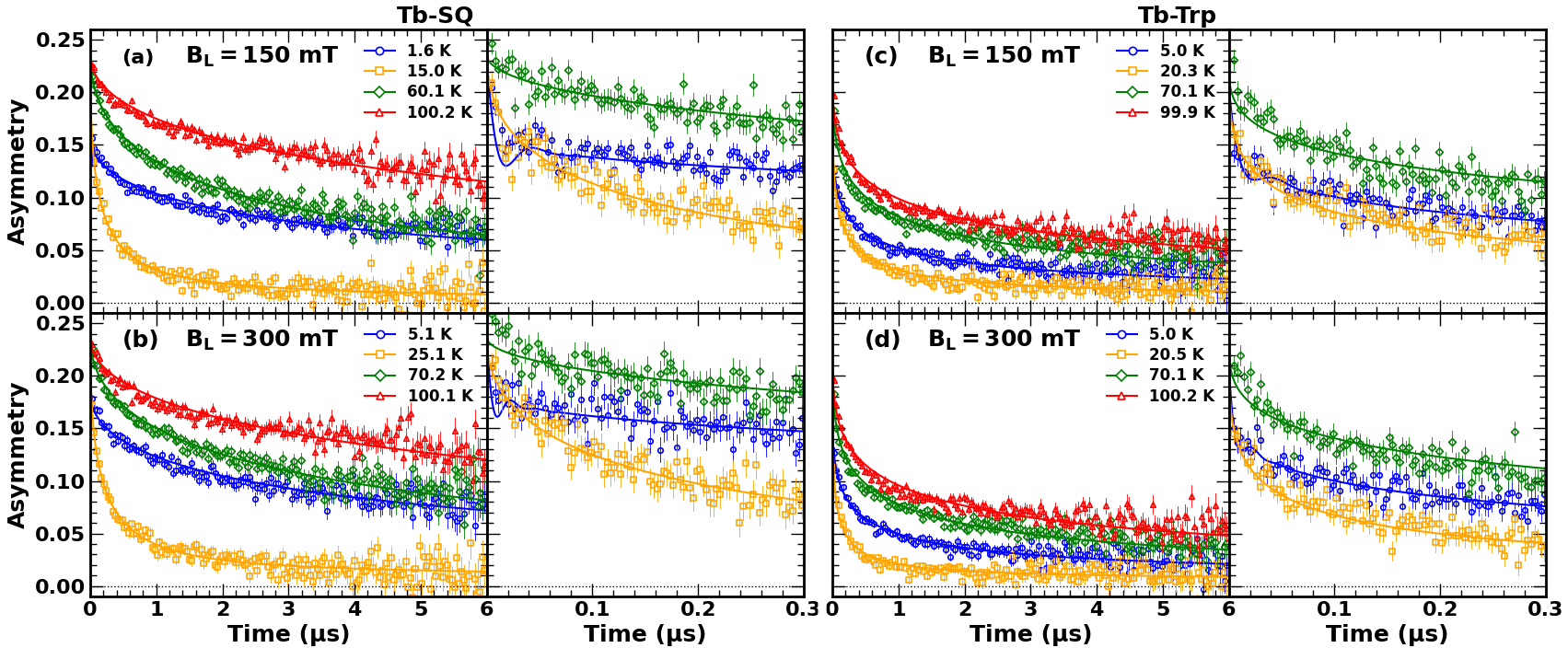} 
    \caption{LF-\muon{} asymmetry time spectra of \tbsq{} sample collected at different representative temperatures measured in applied LF of (a) $B_L= 150~\text{mT}$ and (b) $B_L= 300~\text{mT}$. The solid lines correspond to the best fit obtained with \abbreqn{} \eqref{eqn:FIT_LF_EQN} (see text). The right panel is a zoom for short acquisition time window where selected temperatures are displayed for sake of clarity. The analogous results for \tbtrp{} are shown in the panels (c) and (d), respectively.
    }
    \label{fig:TbSQTbtrp_vtemp_asymmetry_musrfit_results}
\end{figure*}
\subsection{\texorpdfstring{\muon}{}}
\label{sec:results_musr}
In figure \ref{fig:TbSQTbtrp_vtemp_asymmetry_musrfit_results} we report some examples of the time dependence asymmetry spectra measured at representative temperatures for both \tbsq{} and \tbtrp{} samples in LF applied field $B_L=150$ and 300 mT; the long and short time regions are shown in the left and right panels, respectively.  The best phenomenological fitting function of the LF-\muon{} relaxation curves requires two components, with amplitudes $A_1$ and $A_2$, reflecting the existence of two groups of inequivalent muon implantation sites:

\begin{eqnarray}
    \label{eqn:FIT_LF_EQN}
   A(t)  &=&  A_0 \mathcal{P}_{\mathrm{LF}}(t) \nonumber \\
         &=&  A_1 G^{\mathrm{Lor}}_{\mathrm{s}}(\Delta, B_L, t)\exp\big[-(\lambda_1 t)^{\beta}\big] \nonumber\\ & &  + A_2 \exp\big(-\lambda_2 t\big),
\end{eqnarray}

\noindent
{where $A_0$ is the total muon asymmetry calibrated at high temperature and $\mathcal{P}_{\mathrm{LF}}$ is the muon spin polarisation under LF configuration. 
$A_1$ accounts for muon sites closer to the magnetic ions which at low temperatures probes a disordered distribution of quasi-static local fields described by the LF static Lorentzian Kubo-Toyabe (KT) function $G^{\mathrm{Lor}}_{\mathrm{s}}$ \cite{PhysRevB.31.546}:

\begin{eqnarray}
    \label{eqn:FIT_LFKT}
    G^{\mathrm{Lor}}_{\mathrm{s}} &=&  1-\left(\frac{\Delta}{\omega_L}\right)j_1(\omega_L t)\exp{\left(-\Delta t\right)} \nonumber \\
     & & -\left(\frac{\Delta}{\omega_L}\right)^2 \left[j_0(\omega_L t)\exp{\left(-\Delta t\right)}-1\right] \nonumber \\
    & & -\Delta\left[1+\left(\frac{\Delta}{\omega_L}\right)^2\right]\int^t_0 \exp{\left(-\Delta\tau\right)}j_0(\omega_L\tau)\mathrm{d}\tau, \quad
\end{eqnarray}

\noindent
where $\Delta/\gamma^{\mu}$ is the Lorentzian half-width at half-maximum related to the local field distribution width, $\omega_L = \gamma^{\mu} B_L$ is the Larmor frequency at the applied $B_L$ field, $j_0$ and $j_1$ are the zeroth and first-order Bessel functions of the first kind, respectively. {The fit requires that the KT function is multiplied by a stretched-exponential, $\exp\big[-(\lambda_1 t)^{\beta}\big]$, indicating that an independent fluctuating phenomenon with a mean relaxation rate $\lambda_1$ is superimposed \cite{C8CC04703K, PhysRevB.97.144414, PhysRevB.82.174427, doi:10.1021/nn3031673} The stretched behavior is typically observed in presence of an inhomogeneous distribution of local relaxation rates, $\lambda_{\mathrm{loc}}$ \cite{PhysRevLett.77.1386, PhysRevB.74.184430}, with $\beta \le 1$, which means a quasi-random distribution of several non-equivalent implantation sites (for muons closer to the magnetic ions), an occurrence due to many electric potential wells in the compounds and already observed in many molecular magnets \cite{Borsa2006, PhysRevB.97.144414, PhysRevB.100.174416, PhysRevB.102.195424}}.

The component characterized by $A_2$ amplitude accounts for a small ($<$10\% of $A_0$) fraction of muons which are weakly coupled to the magnetic moments, and it is described in the whole $T$ range by a simple exponential decay function with a slow relaxation rate, $\lambda_2\ll\lambda_1$. The fits of \abbreqn{} \eqref{eqn:FIT_LF_EQN} to the data were obtained by global fitting via least-squares optimization using MUSRFIT \cite{SUTER201269}. The fit curves are displayed by the solid lines in \abbrfig{} \ref{fig:TbSQTbtrp_vtemp_asymmetry_musrfit_results}.

For \tbsq{} data, the fit to \abbreqn{} \eqref{eqn:FIT_LF_EQN} looks highly satisfactory, well reproducing also the small dip typically reflecting the distribution of the local fields of the order of few tens of milli-Tesla. On the other hand, \tbtrp{} shows a very fast initial relaxation rate $\lambda_1$ which causes a small loss of the initial asymmetry as observed in other Tb-based MNMs \cite{PhysRevB.79.220404, PhysRevB.82.134401}. In addition, here the dip looks less pronounced and the fit to \abbreqn{} \eqref{eqn:FIT_LF_EQN} is less satisfactory at short times, anyway following the whole trend in particular at longer times where the KT tail is damped.

For the fitting routine, the parameters $A_1$, $A_2$ and $\beta$ are calibrated and held constant for each sample, while the main temperature and field dependent fit parameters are $\Delta, \sigma, \lambda_1$ and $\lambda_2$. The $A_2/A_1$ ratio turns out to be $<$$0.1$ for both the samples, while $\beta$ assumes the values 0.5 and $\simeq$0.35 for \tbsq{} and \tbtrp{}, respectively.

The parameters $\lambda_i (i=1,2)$ in \abbreqn{} \eqref{eqn:FIT_LF_EQN} describe the spin-lattice relaxation rate of the muon spins implanted at specific different muon sites. These relaxation rates are strongly inﬂuenced by the \tbIII{} (electron) spin dynamics through the hyperﬁne muon-electron coupling. In the weak collision approach, where the hyperﬁne muon-electron interaction is treated as a perturbation, it can be demonstrated that the ﬁeld expression for the longitudinal relaxation rate $\lambda_i$ is \cite{Borsa2006, 10.1143/PTP.16.23}:

\begin{equation}
\label{eqn:MNMs_EQN}
\lambda_i(T) \propto \chi T \cdot J(\omega_L^{\mu}) 
\end{equation}

\noindent
where $\chi$ is the magnetic susceptibility and $J(\omega_L^{\mu})$ is the spectral density of the electronic spin ﬂuctuations at the Larmor (muon) frequency $\omega_L^{\mu}$.
\begin{figure}
    \centering
    \includegraphics[width=1.0\columnwidth]{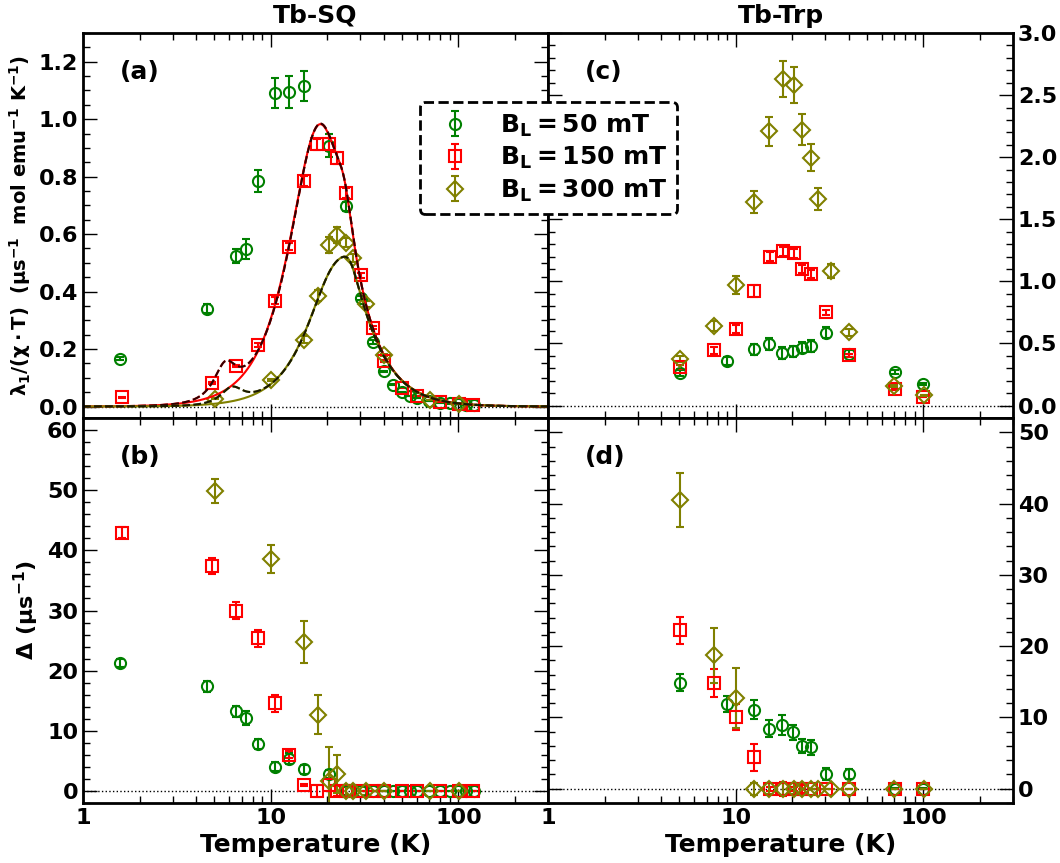}
    \caption{Temperature dependence of (a) relaxation rate $\lambda_1/(\chi T)$, and (b) width of local static magnetic field distribution $\Delta$, obtained from LF-\muon{} measurements on \tbsq{} at applied fields $B_L=$ 50, 150 and 300 mT. The analogous results for Tb-Trp are shown in the panels (c) and (d), respectively. The solid lines represent the results of global curve fitting based on \abbreqn{} \eqref{eqn:BPP_EQN_LT_HT} which include two correlation times $\tau_{c,1}$ and $\tau_{c,2}$, while the dashed (black) lines include an additional correlation time $\tau_{c,3}$ dominating at low temperatures $T \lesssim 8$ K, as described in the main text.
    }
    \label{fig:TbSQTbtrp_vtemp_parameters_musrfit_results}
\end{figure}
Figure \ref{fig:TbSQTbtrp_vtemp_parameters_musrfit_results} displays the temperature dependence of $\lambda_1~(T)/ \chi T$ (where $\lambda_1$ is the larger of the two relaxation rates) and of the field distribution width of the Lorentzian KT, $\Delta(T)$, at applied field $B_L = $ 50, 150 and 300 mT for the two samples. It is worth noting that $\chi T$ values have been taken directly from SQUID experimental data. The $\lambda_2$ relaxation rate is more than one order of magnitude smaller than $\lambda_1$ and follows a similar temperature behavior but, due to the small value of the correspondent muon fraction, the error bars are much larger; for this reason its temperature dependence is not displayed here.

The Lorentzian width $\Delta$ [\abbrfig{} \ref{fig:TbSQTbtrp_vtemp_parameters_musrfit_results}(b) and \ref{fig:TbSQTbtrp_vtemp_parameters_musrfit_results}(d)] shows an increase below $\sim$15 K indicating that spontaneous quasi-static internal fields develop at low temperature.  Approximately at the temperature where this field distribution appears, the quantity $\lambda_1 / \chi T$ [\abbrfig{} \ref{fig:TbSQTbtrp_vtemp_parameters_musrfit_results}(a) and \ref{fig:TbSQTbtrp_vtemp_parameters_musrfit_results}(c)] shows a peak for both the samples. 
As one can see, the $\lambda_1/ \chi T$ peak of the two samples displays an almost opposite  behavior when the applied $B_L$ field is increased: its height decreases and its position shifts toward higher temperatures for \tbsq{}, while  it increases and its position is almost field independent ($\simeq$$20$ K) for \tbtrp{}.

Typically a peak in the behavior of $\lambda_1 /\chi T$ is expected when $\omega_L^{\mu} \tau_c \approx 1, \omega_L^{\mu}$ being the frequency of the measuring probe (muon Larmor frequency $\sim$$10^7\textrm{--}10^8~\text{rad/s}$) and $\tau_c$ the correlation time of the spin dynamics occurring in the system, related to the frequency of the spin ﬂuctuations $\nu_c=1/\tau_c$. In our system, assumed to be composed of non-interacting molecules, the dominating correlation times are thought to be mainly related to the Tb spin flip regulated by a thermally activated spin reversal over the anisotropy barrier, even though a more proper analysis \cite{PhysRevLett.94.077203} involve the transition probabilities of allowed transitions among the different energy levels. Hence a spin slowing dynamics yielding a ``freezing'' process is expected once the low temperature region is approached (slow fluctuation regime with $\tau_c<1/\omega_L^{\mu}$). In this regime, the muons probe a quasi-static ordered lattice of magnetic moments, here evidenced by a field distribution $\Delta >0$. The global $\lambda_1$ vs temperature behavior is generally well described by BPP law \cite{BLOEMBERGEN1947, PhysRev.73.679, PhysRevB.96.184403}:

\begin{equation}
    \label{eqn:BPP_EQN}
    \lambda_1(T,B_L) = C \cdot \chi T \cdot \frac{\tau_c }{1 + \big[\big(\gamma^{\mu} B_L\tau_c\big)^2\big]}
\end{equation}

\noindent
where $C\cdot \chi T \cdot \tau_c = 2(\gamma_{\mu}\delta B_{\mathrm{loc}}^{\mathrm{rms}})^2\tau_c$, $C$ is a scaling constant representing the geometrical part of the hyperfine field fluctuations at the muon site(s), $\delta B_{\mathrm{loc}}^{\mathrm{rms}}$ is the rms width of fluctuating local hyperfine field and $\tau_c$ is temperature-dependent spin-spin correlation time that could follow a power-law \cite{PhysRevB.70.134434, PhysRevB.79.064421} or a thermally activated behavior described by the Arrhenius-like law \cite{PhysRevB.61.R9265, PhysRevLett.85.642, Curro_2009, PhysRevB.88.104503, PhysRevB.92.020505, 10.1063/1.1558597, PhysRevB.14.2005}. The energy barrier obtained from calculating $\lambda_1$ for \tbsq{} (see \abbrfig{} \ref{fig:TbSQTbtrp_vtemp_parameters_musrfit_results}(a)) using the Arrhenius law $\tau_c=\tau_0\exp(\sigma_{\mathrm{A}}/k_B T)$ in \abbreqn{} \eqref{eqn:BPP_EQN} is $\sigma_{\mathrm{A}}/k_B = 66(5)$ K (46 cm$^{-1}$), comparable to the value of 46(1) K derived from AC susceptibility measurements. However, this single correlation time model fails to adequately capture the whole temperature dependence of $\lambda_1$ as a function of $B_L$, particularly at low temperatures. This evidence indicates that other relaxation processes like direct or Raman processes may be needed especially at low temperatures to fully describe the system dynamics \cite{trensal1}.

\begin{table}[htbp!]
    \centering
    \caption{Summary of the global fitting procedures for applied longitudinal fields $B_L$ = 150 and 300 mT for \tbsq{}.
    }
    \begin{ruledtabular}
    \begin{tabular}{lccc}
         & Parameters & Model I & Model II \\ \hline
        \multirow{3}{*}{\textbf{1} Raman} & 
        ${C_1}~(10^{14} ~\text{s}^{-2})$ & $2.50(6)$ & $2.50(6)$ \\ 
        & $c_{0,1}~(10^{-5}~\text{s K}^{\alpha_1})$ & $5(2)$ & $6(2)$ \\ 
        & $\alpha_1$ & $3.0(1)$ & $3.1(1)$ \\ 
        \cmidrule{2-4}
        \multirow{3}{*}{\textbf{2} Orbach} & 
        ${C_2}~(10^{13} ~\text{s}^{-2})$ & $3(0)$ (fixed) & $3(0)$ (fixed) \\ 
        & $\tau_{0,2}~(10^{-13} ~\text{s})$ & $1.0(3)$ & $1.0(5)$ \\ 
        & $\sigma_{A}/k_B~(\text{K})$ & $280(3)$ &  $280(2)$ \\ 
        \cmidrule{2-4}
        \multirow{3}{*}{\textbf{3} Direct} & 
        ${C_3}~(10^{13} ~\text{s}^{-2})$ & $\textrm{--}$ &  $3(1)$ \\ 
        & $c_{0,3}~(\text{s K}^{\alpha_3})$ & $\textrm{--}$ & $0.02(1)$ \\ 
        & $\alpha_3$ & $\textrm{--}$ & $8.5(3)$ \\ 
    \end{tabular}
    \end{ruledtabular}
    \label{tab:SUMMARY_BPP_FIT}
\end{table}

As a consequence, $\lambda_1$ in \abbreqn{} \eqref{eqn:BPP_EQN} is  written as (fitting Model n. I):

\begin{eqnarray}
    \label{eqn:BPP_EQN_LT_HT}
   \lambda_1(T,B_L,\tau_{c, 1},\tau_{c, 2})  &=&   \lambda_1(T,B_L,\tau_{c, 1})_{\mathrm{LT}} \nonumber \\
                                             & &  +\lambda_1(T,B_L,\tau_{c, 2})_{\mathrm{HT}},
\end{eqnarray}

\noindent
where to account for $\lambda_1$ observed at low ($T\lesssim 15K$) and high ($T\gtrsim 15K$) temperatures in the experimental data (LT and HT, respectively),  we assume the following: (\textbf{1}) $\tau_{c, 1} = c_{0, 1} T^{-\alpha_1}$ representing a power-law-like $T$ dependence for correlation time with an exponent $\alpha_1$  at low-temperature \cite{PhysRevB.70.134434}; (\textbf{2})  $\tau_{c, 2} = \tau_{0, 2} \exp(\sigma_{\mathrm{A}}/k_B T)$, representing an Arrhenius-like $T$ dependence for correlation time with energy barrrier $\sigma_{\mathrm{A}}/k_B$ at high-temperature \cite{PhysRevB.61.R9265, PhysRevLett.85.642, Curro_2009, PhysRevB.88.104503, PhysRevB.92.020505, 10.1063/1.1558597, PhysRevB.14.2005}. Thus, the model includes two possible relaxation processes, \textit{i.e.} Raman or Raman-like (\textbf{1}) and Orbach (\textbf{2}) mechanisms. The results of the global fitting using \abbreqn{} \eqref{eqn:BPP_EQN_LT_HT} are shown in \abbrfig{} \ref{fig:TbSQTbtrp_vtemp_parameters_musrfit_results}(a) as continuous lines, and the best-fit parameters summarized in \abbrtab{} \ref{tab:SUMMARY_BPP_FIT} (column Model n. I). As depicted in \abbrfig{} \ref{fig:TbSQTbtrp_vtemp_parameters_musrfit_results}(a), the fitting reproduces the experimental data very well despite the deviation observed for $T \lesssim 8$ K, obtaining the parameters $C_1 = 2.50(6) \times 10^{14}~\text{s}^{-2}$, $c_{0,1} = 5(2) \times 10^{-5}~\text{s K}^{\alpha_1}$, $\alpha_1=3.0(1)$, $C_2 = 3(0) \times 10^{13}~\text{s}^{-2}$, $\tau_{0,2} = 1.0(3) \times 10^{-13}~\text{s}$, and $\sigma_{\mathrm{A}}/k_B=280(2)~\text{K}~(195~\text{cm}^{-1})$. 
Note that \abbreqn{} \eqref{eqn:BPP_EQN_LT_HT} is entirely phenomenological, and we emphasize that the fitting parameters should not be overinterpreted. 
%
%
%
\begin{figure}
    \centering
    \includegraphics[width=\columnwidth]{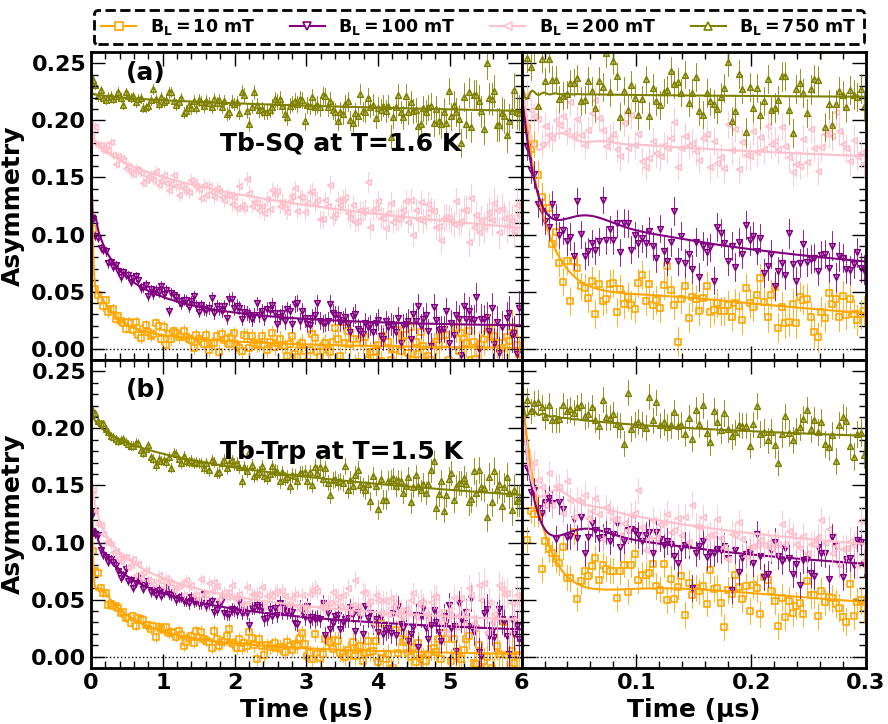}
    \caption{LF-\muon{} asymmetry time spectra at a fixed temperature of (a) $T = 1.6$ K for \tbsq{} and (b) $T = 1.5$ K for \tbtrp{} at representative applied fields $B_L$. 
    The solid lines correspond to the best fit obtained with \abbreqn{} \eqref{eqn:FIT_LF_EQN} (see text). The left and right windows display the long and short-time domains. 
    }
    \label{fig:TbSQTbtrp_field_asymmetry_musrfit_results}
\end{figure}
%
%
%
However, we suggest that in \tbsq{}, two different independent relaxation dynamics are active: the first dominating at low temperature is described as Raman-like \cite{AbragamEPR1970, https://doi.org/10.1002/pssb.2221170202, doi:10.1021/acs.inorgchem.6b00854} and characterized by a correlation time $\tau_{c, 1}$, and the second seen as an Orbach-like process characterized by a correlation time $\tau_{c, 2}$ with energy barrier $\sigma_{\mathrm{A}}/k_B=280~\text{K}~(195~\text{cm}^{-1})$.

The observation of a contribution from an Orbach thermally activated process with an activation barrier of 280 K (195 cm$^{-1}$) for \tbsq{} allows us to make some hypotheses on the relative magnitude of the exchange coupling and ligand field splitting in this system. The zero-field energy levels pattern of \tbsq{} can be described using the following Hamiltonian \cite{rusnati2019thesis}: 

\begin{eqnarray}
\label{eqn:ZF_ELEVELS_HAMILTO}
        \mathcal{H} = \mathcal{H}_{\mathrm{ZFS}} + \mathcal{H}_{\mathrm{exch}} &=&  B_2^0 \big[3J_{z, \mathrm{Tb}}^2 - J_{\mathrm{Tb}}(J_{\mathrm{Tb}}+1)\big] \nonumber\\ 
        & & -2J_{\mathrm{ex}}\bm{J}_{\mathrm{Tb}}\cdot\bm{S_R}
\end{eqnarray}
\noindent
Since to observe Orbach dynamics a strong easy axis anisotropy is required for \tbIII{}, then $B_2^0$ must be large and dominating over the exchange coupling. This provides pair of doublets, identified as $|M_{J,{\mathrm{Tb}}}, M_{S, R}\rangle$, separated by $\sim (2M_{J,{\mathrm{Tb}}} + 1) \cdot J_{\mathrm{ex}}$ within each pair, and by $B_2^0\big[2M_{J,{\mathrm{Tb}}}-1\big]$ between the barycenter of each pair. If we assume, in first approximation, that the spin-spin exchange interaction in \tbsq{} is the same as in the Gadolinium(III) derivative, for which it has been easily determined by DC magnetic measurements \cite{ACaneschi2000}, then $J_{\mathrm{ex}}$ in \abbreqn{} \eqref{eqn:ZF_ELEVELS_HAMILTO} is antiferromagnetic, in agreement with previous studies \cite{B401144A}, and equal to  $-2.9~\text{cm}^{-1}$ ($-4.2~\text{K}$). The ground doublet will then be the one resulting from the antiparallel alignment of the maximal $M_J$ component of the Terbium ion with the spin of the radical $|\pm6 ~\mp~ 1/2 \rangle$ ($|11/2 ~\pm~ 11/2\rangle$ in the coupled representation), whereas in the first excited state the same two components are parallel $|\pm6 ~\pm~ 1/2 \rangle$  (i.e. $|11/2 ~\pm~11/2 \rangle$  in the coupled representation). The observation of an activation barrier of $\sim$300 K for the Orbach process via \muon{} suggests that in the observed temperature range relaxation occurs via the second excited state $|\pm 5 ~\mp~ 1/2 \rangle$  thus providing an estimate for $B_2^0$  of $-6.3~\text{cm}^{-1}$ ($-9.1~\text{K}$). Interestingly, with these assumptions  the first excited state $|\pm6 ~\pm~ 1/2 \rangle$ , is calculated to lie 55 K above the ground state, in reasonable agreement with the barrier derived by AC susceptibility in applied field.

The quality of the global fit of \abbrfig{} \ref{fig:TbSQTbtrp_vtemp_parameters_musrfit_results}(a) using \abbreqn{} \eqref{eqn:BPP_EQN_LT_HT} can be improved significantly (fitting Model n. II) by assuming an additional relaxation process of correlation times $\tau_{c, 3}$ that follows a power law $\tau_{c, 3} = c_{0, 3} T^{-\alpha_3}$. Repeating the global fitting, we obtain the results represented by the dashed lines in \abbrfig{} \ref{fig:TbSQTbtrp_vtemp_parameters_musrfit_results}(a). The fit results are reported in \abbrtab{} \ref{tab:SUMMARY_BPP_FIT} (column Model n. II). it should be noted that our experimental data at low temperatures are not dense enough for displaying a well-defined peak at $T \lesssim 8$ K but, despite this, we obtain a reasonable value $\alpha_3 = 8.5$ very close to the expected $\alpha_3 = 9$  for the Raman process in semi-integer spin systems \cite{Orbach1}.  Thus, our \muon{} suggests an additional relaxation mechanism controlling the spin dynamics of \tbsq{} for $T \lesssim 8$ K.

In general, the results on \tbsq{} show that the BPP-like behavior is reasonably well reproduced, in particular with a good scaling of the peak position and its intensity, that diminishes as $\sim $$1/B_L$, as predicted by the BPP law. A deviation is observed at the lowest field $B_L=50~\text{mT}$, possibly because of the presence of internal fields with magnitude comparable to the externally applied one. Hence, based on this argument we safely conclude not to include this field in the global fitting procedure displayed in \abbrfig{} \ref{fig:TbSQTbtrp_vtemp_parameters_musrfit_results}(a) and described above.

On the contrary, for \tbtrp{} the height of the $\lambda_1 /\chi T$ peak, shown in \abbrfig{} \ref{fig:TbSQTbtrp_vtemp_parameters_musrfit_results}(c), follows an opposite behavior as a function of the applied field. This indicates that the spin dynamics is faster when the external field is increased, fully at odd with a BPP-like behavior. Noteworthy the two molecules are almost isostructural and contain the same Ln magnetic ion, thus pointing to strictly similar ligand field effects. The only main difference is the ligand magnetic character, being paramagnetic for \tbsq{} and diamagnetic for \tbtrp{}. The lack of the BPP behavior here observed very clearly for \tbtrp{}, qualitatively resembles the behavior observed in $\text{Ln}(trensal)$ ($\text{Ln}=$ Er and Dy) SIMs \cite{PhysRevB.100.174416}, in which the ligand is diamagnetic and the strongly mixed nature of the ground doublet results in a spin dynamics dominated by non-Orbach relaxation. This indicates that the intramolecular exchange interaction (and its absence in the derivative with the diamagnetic ligand) has dramatic effects on the spin dynamics.

As a next step we then investigated the field dependence of muon relaxation rate as a function of the applied LF $B_L$ for the two samples at low temperature ($T\le1.6$ K), to study the role of static and dynamic local internal fields. By applying a field $B_L$ parallel to the initial muon-spin polarization, as the field is increased over a certain limit the muons decouple from the static and dynamic spin fluctuations and hence the muon relaxation rate is expected to decrease. At high enough $B_L$, the Lorentzian KT function $G^{\mathrm{Lor}}_{\mathrm{s}}(\Delta, B_L, t)=1$ is independent of $\Delta$; thus $\Delta$ was fixed in \abbreqn{} \eqref{eqn:FIT_LF_EQN} to the values determined at low $B_L$. Figure \ref{fig:TbSQTbtrp_field_asymmetry_musrfit_results}(a) and \ref{fig:TbSQTbtrp_field_asymmetry_musrfit_results}(b) shows representative asymmetry spectra for \tbsq{} ($T = 1.6$ K) and \tbtrp{}  ($T = 1.5$ K), respectively, in the range $5\le B_L\le750$ mT. The curves are reasonably fitted using  \abbreqn{} \eqref{eqn:FIT_LF_EQN}. It is clear from the asymmetry spectra in \abbrfig{} \ref{fig:TbSQTbtrp_field_asymmetry_musrfit_results}(a) that a $B_L$ of about $\sim$750 mT is sufficient for almost completely suppressing the muon relaxation at low temperature for \tbsq{}. However, there is no clear indication of suppression for \tbtrp{} muon relaxation, even with an applied $B_L$ of 750 mT, suggesting that rapidly fluctuating internal fields in the sample remain. 
Figure \ref{fig:TbSQTbtrp_field_lambda_beta_musrfit_results} summarizes the resulting LF-\muon{} values $\lambda_1$ as a function of $B_L$ for both \tbsq{} and \tbtrp{}. The field dependence of $\lambda_1$ for \tbsq{} at lowest temperature of 1.6 K is fitted by using \abbreqn{} \eqref{eqn:BPP_EQN} assuming the expression

\begin{eqnarray}
    \label{eqn:REDFIELD_EQN}
    \lambda_1(B_L) &=& \chi T  \cdot \Bigg[{C_3} \cdot \frac{\tau_{c,3} }{1 + \big[\big(\gamma^{\mu} B_L\tau_{c,3}\big)^2\big]}  \nonumber\\
    & & {C^{\prime}} \cdot \frac{\tau_{c}^{\prime} }{1 + \big[\big(\gamma^{\mu} B_L\tau_{c}^{\prime}\big)^2\big]} \Bigg],
\end{eqnarray}

\noindent
where $\tau_{c, 3} = c_{0, 3} T^{-\alpha_3}$ and $\tau_{c}^{\prime} = {(1+ K_2 B_L^2)}/{K_1}$. In this way the free parameters for fitting $\lambda_1(B_L)$ are $C^{\prime}$, $K_1$ and $K_2$, while $C_3$, $c_{0, 3}$ and $\alpha_3$ acts as fixed values reported above in \abbrtab{} \ref{tab:SUMMARY_BPP_FIT}; the product of $\chi T$ was assumed to be unity. 
The estimated values of the best-fit curve yield $C^{\prime} = 5.3(5) \times 10^{14} ~\text{s}^{-2}$, $K_1 = 1.0(1) \times 10^8 ~\text{s}^{-1}$ and $K_2 = 132(5) ~\text{Tesla}^{-2}$. It has to be noted that the correlation time of the second term in \abbreqn{} \eqref{eqn:REDFIELD_EQN} is explained by a quantum tunneling process in \abbrref{} \cite{trensal1}. 

\begin{figure}
    \centering
    \includegraphics[width=\columnwidth]{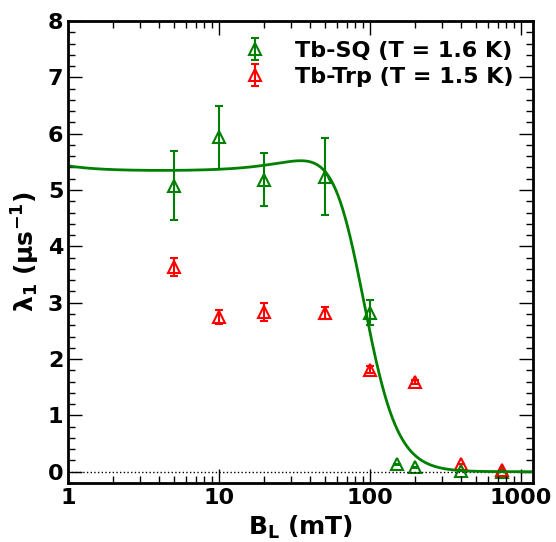}\caption{Longitudinal field dependence of the muon spin relaxation rate $\lambda_1$. The solid line is the best fit obtained with \abbreqn{} \eqref{eqn:REDFIELD_EQN}.
    }
    \label{fig:TbSQTbtrp_field_lambda_beta_musrfit_results}
\end{figure}


\section{\label{sec:conclu} Conclusions}
We have studied the spin dynamics of two Tb-based molecular nanomagnets, isostructural in their first coordination sphere, using \muon{} experiments and AC susceptibility data. The \muon{} data single out completely different spin dynamics in \tbsq{}, where an exchange interaction among the spin of the single \tbIII{} ion and the spin $s=1/2$ of the radical (contained in the SQ ligand) is present, with respect to \tbtrp{}, where the Trp ligand is diamagnetic. The $\lambda_1(T, B_L)$ BPP law, is obeyed by \tbsq{} data once a combination of direct, Raman and Orbach processes with three different correlation times $\tau_{c,i}$ ($i=1,2,3$) is assumed; additionally, at low temperature $T\sim1.5$ K from the \tbsq{} data vs field we single out a contribution to the muon relaxation rate that originates from the quantum tunneling of the magnetization. On the other hand, the BPP law is completely unfulfilled by \tbtrp{} data, where the peaks in the curves collected at different applied fields do not shift with temperature and increase their intensities by increasing the field. 

Thus, it can be concluded that the spin dynamics is highly affected by the magnetic character of the organic ligand: when this is in a radical state, the combined effect of the exchange interaction and ligand field effects result in a magnetic anisotropy barrier to the relaxation, which we suggest occur via the second excited state on the timescale of muon experiments. On the other hand,  the suppression of the exchange interaction channel for relaxation gives rise to a phenomenology which is fully incompatible with the usual dynamic laws applied to the interpretation of \muon{} data of molecular nanomagnets. Interestingly, the observed behaviour for the non-radical system \tbtrp{} is largely resembling the one recently reported by some of us for other non-radical Lanthanide molecular systems \cite{PhysRevB.100.174416}, suggesting that such behavior is more widespread than previously expected. In this sense, the results reported here should stimulate the development of new theoretical models to properly explain the fundamental mechanism here at play.



\section{\label{sec:acknowledgement}Acknowledgement}
This work was financially supported by INFN project NAMASSTE. The PNRR-MUR project PE0000023-NQSTI is also acknowledged. We gratefully thank PSI staff for assistance. 
S.S. was partially supported by the National Centre for HPC, Big Data and Quantum Computing.
%

%
\bibliographystyle{apsrev4-2} 
\bibliography{main}%

\end{document}